\documentclass[12pt]{amsart}
\usepackage{a4wide,color}
\allowdisplaybreaks

\let\pa\partial

\let\eps\varepsilon

\newcommand{\R}{{\mathbb R}}
\newcommand{\T}{{\mathbb T}}

\let\de=\delta
\let\eps=\varepsilon

\let\la=\lambda

\let\pa=\partial

\let\Om=\Omega

\newtheorem{theorem}{Theorem}
\newtheorem{lemma}[theorem]{Lemma}
\newtheorem{proposition}[theorem]{Proposition}

\begin{document}

\title[Boltzmann equation]{Pointwise Behavior of the Linearized Boltzmann Equation on Torus}

\author[K.-C. Wu]{Kung-Chien Wu}
\address{Department of Pure Mathematics and Mathematical Statistics,
University of Cambridge, Wilberforce Road, Cambridge CB3 0WB, United Kingdom}
\email{kcw28@dpmms.cam.ac.uk; kungchienwu@gmail.com}

\date{\today}

\thanks{This paper is supervised by Cl\'{e}ment
Mouhot in Cambridge. It is a pleasure to thank I-Kun Chen, Seung-Yeal Ha, Chanwoo Kim, Hung-Wen Kuo, Chi-Kun Lin, Tai-Ping Liu and Se Eun Noh for stimulating discussion concerning this
paper. This work is supported by the
Tsz-Tza Foundation in institute of mathematics, Academia Sinica,
Taipei, Taiwan. Part of this work was written during the stay
at Department of Mathematics, Stanford University; the author thanks Tai-Ping Liu for his kind hospitality.}

\begin{abstract}
We study the pointwise behavior of the linearized Boltzmann equation on torus
for non-smooth initial perturbation.
The result reveals both the fluid and kinetic aspects of this model. The fluid-like waves are
constructed as part of the long-wave expansion in the spectrum of the Fourier
mode for the space variable, the time decay rate of the fluid-like waves depends on the size of the domain. We design a
Picard-type iteration for constructing the increasingly regular kinetic-like waves, which are carried by the transport equations and have exponential time decay rate. Moreover, the mixture lemma plays an important role in constructing the kinetic-like waves, we supply a new proof of this lemma to avoid constructing explicit solution of the damped transport equations (compare Liu-Yu's proof \cite{[Kuo], [LiuYu]}).
\end{abstract}



\maketitle


\section{Introduction}

The Boltzmann equation for the hard sphere model reads
\begin{equation}\label{bot.1.a}
\left\{\begin{array}{l}
\displaystyle \pa_{t}F+\xi\cdot\nabla_{x}F=\frac{1}{\eps}Q(F,F)\,,
\\ \\
\displaystyle F(x,0,\xi)=F_{0}(x,\xi)\,,
\end{array}
\right.\end{equation}
where
\begin{equation*}
\left\{\begin{array}{l}
\displaystyle Q(g,h)=\frac{1}{2}\int_{U}\big[-g(\xi) h(\xi_{*})-g(\xi_{*})h(\xi)+
g(\xi')h(\xi'_{*})+g(\xi'_{*})h(\xi')\big]|(\xi-\xi_{*})\cdot\Om| d\xi_{*}d\Om\,,
\\ \\
\displaystyle U=\big\{(\xi_{*},\Om)\in\R^{3}\times \mathbb{S}^{2}: (\xi-\xi_{*})\cdot\Om\geq 0\big\}\,,
 \\ \\
\xi'=\xi-[(\xi-\xi_{*})\cdot\Om]\Om\,,\quad \xi'_{*}=\xi+[(\xi-\xi_{*})\cdot\Om]\Om\,.
\end{array}
\right.\end{equation*}
Here $\eps$ is the Knudsen number, the microscopic velocity $\xi\in\R^{3}$ and the space variable $x\in \T^{3}_{1}$, the 3-dimensional torus with unit size of each side. In order to remove the parameter $\eps$ from the equation, we introduce the new scaled variables:
$$
\widetilde{x}=\frac{1}{\eps}x\,, \quad \widetilde{t}=\frac{1}{\eps}t\,,
$$
then after dropping the tilde, the equation (\ref{bot.1.a}) becomes
\begin{equation}\label{bot.1.b}
\left\{\begin{array}{l}
\displaystyle \pa_{t}F+\xi\cdot\nabla_{x}F=Q(F,F)\,,\quad (x,t,\xi)\in \T^{3}_{1/\eps}\times\R^{+}\times\R^{3}\,,
\\ \\
\displaystyle F(x,0,\xi)=F_{0}(x,\xi)\,,
\end{array}
\right.\end{equation}
where $\T^{3}_{1/\eps}$ denotes the 3-dimensional torus with size $1/\eps$ of each side.
The conservation laws of mass, momentum, as well as energy, can be formulated as
\begin{equation}\label{bot.1.c}
\frac{d}{dt}\int_{\T^{3}_{1/\eps}}\int_{\R^{3}}\Big\{1,\xi,|\xi|^{2}\Big\}F(t,x,\xi)d\xi dx=0\,.
\end{equation}
It is well-know that Maxwellians are steady states to the Boltzmasnn equation. Thus, it is natural to linearize the Boltzmann equation (\ref{bot.1.b}) around a global Maxwellian
$$
w(\xi)=\frac{1}{(2\pi)^{3/2}}\exp\Big(\frac{-|\xi|^{2}}{2}\Big)\,,
$$
with the standard perturbation $f(t,x,\xi)$ to $w$ as
$$
F=w+w^{1/2}f\,.
$$
Then after substituting into (\ref{bot.1.b}) and dropping the nonlinear term, we have the linearized Boltzmann equation
\begin{equation}\label{bot.1.d}
\left\{\begin{array}{l}
\displaystyle \pa_{t}f+\xi\cdot\nabla_{x}f=2Q(w,w^{1/2}f)=Lf\,,
\\ \\
\displaystyle f(x,0,\xi)=I(x,\xi)\,,
\end{array}
\right.\end{equation}
here we define $f(x,t,\xi)=\mathbb{G}_{\eps}^{t}I(x,\xi)$, i.e. $\mathbb{G}_{\eps}^{t}$ is the solution operator (Green function) of the linearized Boltzmann equation (\ref{bot.1.d}). Assuming the initial density distribution function $F_{0}(x,\xi)$ has the same mass, momentum and total energy as the Maxwellian $w$, we can further rewrite the conservation laws (\ref{bot.1.c}) as
\begin{equation}\label{bot.1.e}
\int_{\T^{3}_{1/\eps}}\int_{\R^{3}}w^{1/2}(\xi)\Big\{1,\xi,|\xi|^{2}\Big\}I(x,\xi)d\xi dx=0\,.
\end{equation}
This means that the initial condition $I$ satisfies the zero mean condition.

Before the presentation of the properties of the collision operator $L$, let us define some notations in this paper.
For microscopic variable $\xi$, we shall use $L^{2}_{\xi}$
to denote the classical Hilbert space with norm
$$\Vert
f\Vert_{L^{2}_{\xi}}=\Big(\int_{\R^{3}} |f|^2d\xi\Big)^{1/2}\,,$$
the Sobolev space of functions with all its $s$-th partial derivatives in $L^{2}_{\xi}$
will be denoted by $H^{s}_{\xi}$. The $L^{2}_{\xi}$ inner product in $\R^{3}$ will be denoted by $\big<\cdot,\cdot\big>_{\xi}$. We denote the weighted sup norm as
$$
\|f\|_{L^{\infty}_{\xi,\beta}}=\sup_{\xi\in \R^{3}}|f(\xi)|(1+|\xi|)^{\beta}\,.
$$
For space variable $x$, we shall use $L^{2}_{x}$
to denote the classical Hilbert space with norm
$$\|f\|_{L^{2}_{x}}=\Big(\frac{1}{|\T^{3}_{1/\eps}|}\int_{\T^{3}_{1/\eps}}|f|^{2}dx\Big)^{1/2}\,,$$
the Sobolev space of functions with all its $s$-th partial derivatives in $L^{2}_{x}$
will be denoted by $H^{s}_{x}$. We denote the sup norm as
$$
\|f\|_{L^{\infty}_{x}}=\sup_{x\in \T^{3}_{1/\eps}}|f(x)|\,.
$$
\begin{proposition}{\rm(\cite{[Cercignani]})}\label{pro1} The collision operator $L$ consists of a multiplicative operator $\nu(\xi)$ and an integral operator $K$:
$$
Lf=-\nu(\xi)f+Kf\,,
$$
where
\begin{equation*}
\left\{\begin{array}{l}
\displaystyle Kf=\int_{\R^{3}}W(\xi,\xi_{*})f(\xi_{*})d\xi_{*}\,,
\\ \\
W(\xi,\xi_{*})=\frac{2}{\sqrt{2\pi}|\xi-\xi_{*}|}\exp\big\{-\frac{(|\xi|^{2}-|\xi_{*}|^{2})^{2}}{8|\xi-\xi_{*}|^{2}}
-\frac{|\xi-\xi_{*}|^{2}}{8}\big\}
-\frac{|\xi-\xi_{*}|}{2}\exp\big\{-\frac{|\xi|^{2}+|\xi_{*}|^{2}}{4}\big\}\,,
\\ \\
\nu(\xi)=\frac{1}{\sqrt{2\pi}}\Big[2e^{-\frac{|\xi|^{2}}{2}}+\big(|\xi|+|\xi|^{-1}\big)\int_{0}^{|\xi|}e^{-\frac{u^{2}}{2}}du\Big]\,.
\\
\end{array}
\right.\end{equation*}
For multiplicative operator $\nu(\xi)$, there exists positive lower bound $\nu_{0}$ such that $\nu(\xi)\geq \nu_{0}$ for all $\xi\in\R^{3}$.
Moreover, the derivatives of $\nu(\xi)$ in $\xi$ is bound, i.e. for all multi index $\alpha$,
\begin{equation}\label{bot.1.f}
|\pa^{\alpha}_{\xi}\nu(\xi)|\leq C_{\alpha}\,.
\end{equation}
The integral operator $K$ has smoothing property in $\xi$, i.e.,
\begin{equation}\begin{array}{l} \label{bot.1.g}
\|Kh\|_{H^{1}_{\xi}}\leq C\|h\|_{L^{2}_{\xi}}\,,\quad
\|Kh\|_{L^{\infty}_{\xi,\beta}}\leq C\|h\|_{L^{\infty}_{\xi,\beta}}\,,
\end{array}\end{equation}
for any $\beta\geq 0$.
\end{proposition}

The integral operator $K$ can be decomposed into the singular part and the regular part, i.e., $K=K_{s}+K_{r}$, $K_{s}\equiv K_{s,D}$ and $K_{r}\equiv K_{r,D}$:

\begin{equation*}
\left\{\begin{array}{l}
\displaystyle K_{s}f=\int_{\R^{3}}\chi\Big(\frac{|\xi-\xi_{*}|}{D\nu_{0}}\Big)W(\xi,\xi_{*})f(\xi_{*})d\xi_{*}\,,
\\ \\
K_{r}f=Kf-K_{s}f\,,
\\ \\
\chi(r)=1 \quad \hbox{for} \quad r\in [-1,1]\,,
\\ \\
{\rm supp}(\chi)\subset [-2,2]\,, \quad \chi\in C^{\infty}_{c}(\R)\,, \quad  \chi\geq 0\,.
\\
\end{array}
\right.\end{equation*}
\begin{proposition}\label{pro2}{\rm(\cite{[LiuYu]})}
The singular part $K_{r}$ shares the same smoothing properties as $K$ and has strength of the order of the cut-off parameter $D$:
\begin{equation}\label{bot.1.h}
\|K_{s}h\|_{H^{1}_{\xi}}\leq D\|h\|_{L^{2}_{\xi}}\,,\quad
\|K_{s}h\|_{L^{\infty}_{\xi,\beta}}\leq D\|h\|_{L^{\infty}_{\xi,\beta}}\,.
\end{equation}
The regular part $K_{r}$ has better smoothing property in $\xi$: for all $s>0$,
\begin{equation}\label{bot.1.i}
\|K_{r}h\|_{H^{s}_{\xi}}\leq C\|h\|_{L^{2}_{\xi}}\,.
\end{equation}
\end{proposition}
In order to estimate the Green function of the linearized Boltzmann equation
 in next section, we need to recall the spectrum ${\rm Spec}(\eps k)$, $k\in \mathbb{Z}^{3}$, of the operator $-i\pi\eps\xi\cdot k +L$
\begin{proposition}{\rm(\cite{[Ellis]})}\label{pro3}
There exists $\de>0$ and $\tau=\tau(\de)>0$ such that
\\
\noindent{\rm(i)}\quad  For any $|\eps k|>\de$,
\begin{equation}\label{bot.1.j}
\hbox{\rm Spec}(\eps k)\subset\{z\in \mathbb{C} : {\rm Re}(z)<-\tau\}\,.
\end{equation}
\\
\noindent{\rm(ii)}\quad  For any $|\eps k|<\de$, the spectrum within the region $\{z\in\mathbb{C} : Re(z)>-\tau\}$ consisting of exactly five eigenvalues $\{\sigma_{j}(\eps k )\}_{j=0}^{4}$,
\begin{equation}\label{bot.1.k}
\hbox{\rm Spec}(\eps k)\cap\{z\in\mathbb{C} : {\rm Re}(z)>-\tau\}=
\{\sigma_{j}(\eps k )\}_{j=0}^{4}\,,
\end{equation}
and the corresponding eigenvectors $\{e_{j}(\eps k )\}_{j=0}^{4}$,
where
\begin{equation*}\begin{array}{l}\label{bot.1.l}
\displaystyle \sigma_{j}(\eps k )=\sum_{n=1}^{3}a_{j,n}(i|\eps k|)^{n}+O(\eps k )^{4}\,,\quad a_{j,2}>0\,,
\\
\displaystyle e_{j}(\eps k )=\sum_{n=1}^{3}e_{j,n}(i|\eps k|)^{n}+O(\eps k )^{4}\,,\quad \big<e_{j},e_{k}\big>_{\xi}=\de_{jk}.
\end{array}
\end{equation*}
\\
\noindent{\rm(iii)}\quad
\begin{equation*}\label{bot.1.m}
\displaystyle e^{(-i\pi\eps\xi\cdot k +L)t}f=\Pi_{\de}f+\chi_{\{|\eps k|<\de\}}\sum_{j=0}^{4}e^{\sigma_{j}(-\eps k)t}\big<e_{j}(\eps k ), f\big>_{\xi}e_{j}(\eps k )
\end{equation*}
where $\|\Pi_{\de}\|_{L^{2}_{\xi}}=O(1)e^{-a(\tau)t}$, $a(\tau)>0$, $\chi_{\{\cdot\}}$ is the indicator function.
\end{proposition}
The fluid behavior is studied by constructing the Green function represented as the Fourier series in space variable $x$:
$$
\mathbb{G}_{\eps}^{t}=\sum_{k\in\mathbb{Z}}\frac{1}{|\T^{3}_{1/\eps}|}e^{i\pi\eps k\cdot x+(-i\pi\eps\xi\cdot k +L)t}\,,
$$
where the Green function is served as a function of $\xi$. The analysis of the Green function is equivalent to the analysis of the spectrum of the operator $-i\pi\eps\xi\cdot k +L$. Notice that the spectrum includes five curves which bifurcate from the origin. The origin is the multiple zero eigenvalues of $L$, the operator at $k=0$. The kernel of $L$ are the fluid variables and the fluid-like waves are constructed from these curves near the origin.

The kinetic aspect of the solution is described by the damped transport equation:
$$
\pa_{t}g+\xi\cdot\nabla_{x}g+\nu(\xi)g=0\,.
$$
The operator $K$ is a smooth operator in $\xi$ variable, we will use this smooth property to design a
Picard-type iteration for constructing the increasingly regular kinetic-like waves.

Once the kinetic-like waves and fluid-like waves were constructed, the rest of the solution is sufficiently smooth and it has exponential time decay rate.
\begin{theorem}\label{theorem1}
Given $\beta>\frac{3}{2}$, for any $I\in L_{\xi,\beta}^{\infty}$ with compact support in $x$ and satisfying the zero mean conditions {\rm(\ref{bot.1.e})}, the solution of $(\ref{bot.1.d})$
$$
f=\mathbb{G}_{\eps}^{t}I=\mathbb{G}_{\eps, F}^{t}I+\mathbb{G}_{\eps, K}^{t}I+\mathbb{G}_{\eps, R}^{t}I\,,
$$
consists of the fluid part $\mathbb{G}_{\eps, F}^{t}I$: smooth in space variable $x$ and the time decay rate depends on the size of the domain, i.e. there exist $\de, \de_{0}, C_{0}>0$ such that for all $s>0$
\\ \\
\noindent{\rm(i)} If $\eps>\de$,
\begin{equation*}
\|\mathbb{G}_{\eps, F}^{t}I\|_{H^{s}_{x}L^{2}_{\xi}}=0\,,
\end{equation*}
\\
\noindent{\rm(ii)} If $\de_{0}<\eps<\de$,
\begin{equation*}
\|\mathbb{G}_{\eps, F}^{t}I\|_{H^{s}_{x}L^{2}_{\xi}}=O(1)e^{-O(1)\eps^{2}t}\,,
\end{equation*}
\\
\noindent{\rm(iii)} If $0<\eps<\de_{0}$,
\begin{equation*}
\|\mathbb{G}_{\eps, F}^{t}I\|_{H^{s}_{x}L^{2}_{\xi}}=O(1)\frac{C_{0}}{(1+t)^{3/2}}e^{-O(1)\eps^{2}t}\,;
\end{equation*}
the kinetic part $\mathbb{G}_{\eps, K}^{t}I$: nonsmooth in space variable $x$ and time decay exponentially
\begin{equation*}
\|\mathbb{G}_{\eps, K}^{t}I\|_{L^{\infty}_{x}L^{\infty}_{\xi,\beta}}=O(1)e^{-O(1)t}\,;
\end{equation*}
and the smooth remainder part $\mathbb{G}_{\eps, R}^{t}I$:
\begin{equation*}
\|\mathbb{G}_{\eps, R}^{t}I\|_{H^{2}_{x}L^{2}_{\xi}}=O(1)e^{-O(1)t}\,.
\end{equation*}
\end{theorem}

The spectrum analysis of the Boltzmann equation was introduced by Ellis-Pinsky \cite{[Ellis]}. Recently, Mouhot \cite{[Mouhot]} gives the explicit coercivity estimates for the linearized Boltzmann operator. The spectrum analysis of the linearized Boltzmann equation has been carried out by many authors. In particular, the exponential time decay rates for the Boltzmann equation with hard potentials on
torus was firstly provided by Ukai \cite{[Ukai1]}. The time-asymptotic nonlinear stability was obtained in \cite{[Nishida3], [Ukai]}. Using Nishida's approach, \cite{[Nishida]} obtained the time-asymptotic
equivalent of Boltzmann solutions and Navier-Stokes solutions.
These works yield
the $L^{2}$ theory, since the Fourier transform is isometric in $L^2$.

The mixture lemma plays an important role in constructing the kinetic-like waves. It states that the mixture of the two operators
$\mathbb{S}$ and $K$ in $\mathbb{M}^{t}_{j}$ (see section \ref{se4} below) transports the regularity in the microscopic velocity
$\xi$ to the regularity of the space and time $(x,t)$. This idea was firstly introduced by Liu-Yu \cite{[LiuYu], [LiuYu2], [LiuYu3], [LiuYu1]} to construct the Green function of the Boltzmann equation. In Liu-Yu's paper \cite{[Kuo], [LiuYu], [LiuYu1]}, the proof of the mixture lemma relies on the explicit solution of the damped transport equations. However, in this paper, we introduce a differential operator to avoid constructing explicit solution, this operator commutes with free transport operator and can transports the microscopic velocity regularity to space regularity, this idea will help us to consider more complicated problems, such as the Fokker-Planck equation or the Landau equation. Recently, we can apply coercivity estimates \cite{[Guo], [Mouhot]} to prove mixture lemma for Landau equation on soft potential \cite{[Wu]}. Actually, the mixture lemma is similar in
spirit to the well-known Averaging Lemma, see \cite{[Bouchut], [Golse], [Jabin]}. These
two lemmas have been introduced independently and used for different purposes.

The pointwise description of the one-dimensional linearized Boltzmann equation with hard sphere was firstly provided by Liu-Yu \cite{[LiuYu]}, the fluid-like waves can be constructed by both complex and spectrum analysis, it reveals the dissipative behavior of the type of the Navier-Stokes equation as usually seems by the Chapman-Enskog expansion. The kinetic-like waves can be constructed by Picard-type iteration and mixture lemma. In this paper, we apply similar ideas on torus, we can also construct the kinetic-like waves and fluid-like waves, which are both time decay exponentially. Moreover, the decay rate of the fluid-like waves depend on the size of the domain.

The rest of the paper is organized as follows. In section \ref{se2}, we construct the Green function of the linearized Boltzmann equation on torus. We use the long wave short wave decomposition and the spectrum analysis to obtain time decay rate. In section \ref{se3}, we improve the estimate of the fluid-like waves. In section \ref{se4}, we design a
Picard-type iteration for constructing the increasingly regular kinetic-like waves. Finally, we supply a new proof of the mixture lemma in the appendix.

\section{Long Wave Short Wave Decomposition}\label{se2}

Consider the linearized Boltzmann equation
\begin{equation}\label{bot.2.a}
\left\{\begin{array}{l}\displaystyle\pa_{t}f+\xi\cdot\nabla_{x}f=Lf\,,\quad(x,t,\xi)\in (\T^{3}_{1/\eps},\R^{+},\R^{3})\,,
\\ \\
\displaystyle f(x,0,\xi)=I(x,\xi)\,,
\\
\end{array}
\right.\end{equation}
where $I$ satisfies the zero mean conditions {\rm(\ref{bot.1.e})}.
Hereafter, we will use just one index to denote the 3-dimensional sums with respect to the vector $k=(k_{1},k_{2},k_{3})\in \mathbb{Z}^{3}$, hence we set
$$
\sum_{k\in\mathbb{Z}}=\sum_{(k_{1},k_{2},k_{3})\in\mathbb{Z}^{3}}\,.
$$
Consider the Fourier series of initial condition $I$ in $x$
\begin{equation}\label{bot.2.b}
\left\{\begin{array}{l}
\displaystyle
I(x,\xi)=\sum_{k\in\mathbb{Z}}(\hat{I})_{k}(\xi)e^{i\pi\eps k\cdot x}\,,
\\ \\
\displaystyle(\hat{I})_{k}(\xi)=\frac{1}{|\T^{3}_{1/\eps}|}\int_{\T^{3}_{1/\eps}}I(\cdot,\xi)e^{-i\pi\eps k \cdot x}dx\,,
\\
\end{array}
\right.\end{equation}
rewrite the solution $f(x,t,\xi)$ of (\ref{bot.2.a}) as Fourier series
\begin{equation}\label{bot.2.c}
\left\{\begin{array}{l}
\displaystyle
f(x,t,\xi)=\sum_{k\in\mathbb{Z}}(\hat{f})_{k}(t,\xi)e^{i\pi\eps k\cdot x}\,,
\\ \\
\displaystyle(\hat{f})_{k}(t,\xi)=\frac{1}{|\T^{3}_{1/\eps}|}\int_{\T^{3}_{1/\eps}}f(\cdot,t,\xi)e^{-i\pi\eps k\cdot x}dx\,,
\\
\end{array}
\right.\end{equation}
the Fourier modes of (\ref{bot.2.b})--(\ref{bot.2.c}) satisfy the following equations
\begin{equation*}
\left\{\begin{array}{l}
\displaystyle
\pa_{t}\hat{f}_{k}+i\pi\eps\xi\cdot k \hat{f}_{k}-L\hat{f}_{k}=0\,,
\\ \\
\displaystyle \hat{f}_{k}(0,\xi)=(\hat{I})_{k}\,.
\\
\end{array}
\right.\end{equation*}
Hence
\begin{equation*}
\hat{f}_{k}(t,\xi)=e^{(-i\pi\eps\xi\cdot k +L)t}(\hat{I})_{k}(\xi)\,,
\end{equation*}
the solution of (\ref{bot.2.a}) is given by
\begin{align*}
  f(x,t,\xi)&=\sum_{k\in\mathbb{Z}}e^{i\pi\eps k\cdot x+(-i\pi\eps\xi\cdot k +L)t}(\hat{I})_{k}(\xi)\,, \\
  &=\sum_{k\in\mathbb{Z}}\frac{1}{|\T^{3}_{1/\eps}|}\int_{\T^{3}_{1/\eps}}e^{i\pi\eps k \cdot(x-y)+(-i\pi\eps\xi\cdot k +L)t}
I(y,\xi)dy\,,
\end{align*}
where the Green function $\mathbb{G}_{\eps}^{t}(x,\xi)$ can be expressed as
\begin{equation}\label{bot.2.d}
\displaystyle
\mathbb{G}_{\eps}^{t}(x,\xi)=\sum_{k\in\mathbb{Z}}\frac{1}{|\T^{3}_{1/\eps}|}e^{i\pi\eps k\cdot x+(-i\pi\eps\xi\cdot k +L)t}\,.
\end{equation}
Note that if $\eps\to 0$, the Green function $\mathbb{G}^{t}_{0}(x,\xi)$ becomes an integral
\begin{equation*}
\displaystyle
\mathbb{G}^{t}_{0}(x,\xi)=\int_{\R^{3}}e^{i\eta\cdot x+(-i\pi\xi\cdot\eta+L)t}d\eta\,.
\end{equation*}
We can decompose the Green function (\ref{bot.2.d}) into the long wave part $\mathbb{G}^{t}_{\eps, L}$ and the short wave part $\mathbb{G}^{t}_{\eps, S}$ respectively
\begin{equation}\begin{array}{l}\label{bot.2.e}
\displaystyle
\mathbb{G}^{t}_{\eps, L}(x,\xi)=\sum_{|\eps k|<\de}\frac{1}{|\T^{3}_{1/\eps}|}e^{i\pi\eps k\cdot x+(-i\pi\eps\xi\cdot k +L)t}\,,
\\ \\
\displaystyle
\mathbb{G}^{t}_{\eps, S}(x,\xi)=\sum_{|\eps k|>\de}\frac{1}{|\T^{3}_{1/\eps}|}e^{i\pi\eps k\cdot x+(-i\pi\eps\xi\cdot k +L)t}\,.
\end{array}
\end{equation}
The following long wave short wave analysis relies on spectrum analysis (Proposition \ref{pro3}).
\begin{lemma}{\rm(Short wave $\mathbb{G}^{t}_{\eps, S}$)} For any $s>0$, $I\in H^{s}_{x}L^{2}_{\xi}$, we have
\begin{equation}\begin{array}{l}\label{bot.2.f}
\|\mathbb{G}^{t}_{\eps, S}I\|_{L^{2}_{x}L^{2}_{\xi}}\leq e^{-O(1)t}\|I\|_{L^{2}_{x}L^{2}_{\xi}}\,,
\\ \\
\|\mathbb{G}^{t}_{\eps, S}I\|_{H^{s}_{x}L^{2}_{\xi}}\leq e^{-O(1)t}\|I\|_{H^{s}_{x}L^{2}_{\xi}}\,.
\end{array}
\end{equation}
\end{lemma}
\noindent{\it Proof.} Note that
\begin{equation*}
\mathbb{G}^{t}_{\eps, S}I=\sum_{|\eps k|>\de}e^{(-i\pi\eps\xi\cdot k +L)t}e^{i\pi\eps k\cdot x}\hat{I}_{k}(\xi)\,.
\end{equation*}
For $L^{2}_{x}L^{2}_{\xi}$ estimate, we have
\begin{equation*}
\|\mathbb{G}^{t}_{\eps, S}I\|^{2}_{L^{2}_{x}}=\sum_{|\eps k|>\de}
|e^{(-i\pi\eps\xi\cdot k +L)t}\hat{I}_{k}(\xi)|^{2}\,,
\end{equation*}
and hence by spectrum property (\ref{bot.1.j}), we obtain
\begin{align*}
 \|\mathbb{G}^{t}_{\eps, S}I\|^{2}_{L^{2}_{x}L^{2}_{\xi}}&=\sum_{|\eps k|>\de}
\|e^{(-i\pi\eps\xi\cdot k +L)t}\hat{I}_{k}(\xi)\|_{L^{2}_{\xi}}^{2}\leq e^{-O(1)t}
\sum_{|\eps k|>\de}
\|\hat{I}_{k}(\xi)\|_{L^{2}_{\xi}}^{2} \\
  &=e^{-O(1)t}\int_{\R^{3}}\sum_{|\eps k|>\de}|\hat{I}_{k}(\xi)|^{2}d\xi
\leq  e^{-O(1)t}\|I\|_{L^{2}_{x}L^{2}_{\xi}}^{2}\,.
\end{align*}
For the high order estimate, we have
\begin{equation*}\begin{array}{l}
\displaystyle \|\mathbb{G}^{t}_{\eps, S}I\|^{2}_{H^{s}_{x}}\leq\sum_{|\eps k|>\de}(1+|\pi\eps k|^{2} )^{s}
|e^{(-i\pi\eps\xi\cdot k +L)t}\hat{I}_{k}(\xi)|^{2}\,,
\end{array}\end{equation*}
and hence
\begin{equation*}\begin{array}{l}
\displaystyle \|\mathbb{G}^{t}_{\eps, S}I\|^{2}_{H^{s}_{x}L^{2}_{\xi}}\leq e^{-O(1)t}\sum_{|\eps k|>\de}
(1+|\pi\eps k|^{2} )^{s}\|\hat{I}_{k}\|_{L^{2}_{\xi}}^{2}
\leq e^{-O(1)t}\|I\|_{H^{s}_{x}L^{2}_{\xi}}^{2}\,.
\end{array}\end{equation*}
\qed

In order to study the long wave part $\mathbb{G}_{\eps, L}^{t}$, we need to decompose the long wave part as the fluid part and non-fluid part, i.e. $\mathbb{G}_{\eps, L}^{t}=\mathbb{G}_{\eps, F}^{t}+\mathbb{G}_{\eps, L;\perp}^{t}$, where
\begin{equation}\begin{array}{l}\label{bot.2.g}
\displaystyle \mathbb{G}_{\eps, F}I=\sum_{|\eps k|<\de}\sum_{j=0}^{4}e^{\sigma_{j}(-\eps k)t}e^{i\pi\eps k\cdot x}\big<e_{j}(\eps k ), \hat{I}_{k}\big>_{\xi}e_{j}(\eps k )\,,
\\ \\
\displaystyle \mathbb{G}^{t}_{\eps, L;\perp}I=\sum_{|\eps k|<\de}e^{i\pi\eps k\cdot x}\Pi_{\de}\hat{I}_{k}\,.
\end{array}\end{equation}
\begin{lemma}{\rm(Long wave $\mathbb{G}^{t}_{\eps, L}$)} For any $s>0$, $I\in L^{2}_{x}L^{2}_{\xi}$ and satisfying the zero mean condition {\rm(\ref{bot.1.e})}, we have
\begin{equation}\begin{array}{l}\label{bot.2.h}
\|\mathbb{G}^{t}_{\eps, L;\perp}I\|_{H^{s}_{x}L^{2}_{\xi}}\leq e^{-O(1)t}\|I\|_{L^{2}_{x}L^{2}_{\xi}}\,,
\\ \\
\|\mathbb{G}_{\eps, F}I\|_{H^{s}_{x}L^{2}_{\xi}}\leq e^{-O(1)\eps^{2} t}\|I\|_{L^{2}_{x}L^{2}_{\xi}}\,.
\end{array}\end{equation}
\end{lemma}
\noindent{\it Proof.} For the non-fluid part, using the spectrum property (\ref{bot.1.j}), we have
\begin{equation*}\begin{array}{l}
\displaystyle\|\mathbb{G}^{t}_{\eps, L;\perp}I\|^{2}_{H^{s}_{x}}\leq\sum_{|\eps k|<\de}\big(1+|\pi\eps k |^{2}\big)^{s}
|\Pi_{\de}\hat{I}_{k}(\xi)|^{2}
\leq\big(1+|\pi\de |^{2}\big)^{s}\sum_{|\eps k|<\de}
|\Pi_{\de}\hat{I}_{k}(\xi)|^{2}\,,
\end{array}\end{equation*}
and hence
\begin{equation*}\begin{array}{l}
\|\mathbb{G}^{t}_{\eps, L;\perp}I\|^{2}_{H^{s}_{x}L^{2}_{\xi}}
\leq e^{-O(1)t}\|I\|_{L^{2}_{x}L^{2}_{\xi}}^{2}\,.
\end{array}\end{equation*}
For the fluid part, by (\ref{bot.1.k}) and the zero mean conditions {\rm(\ref{bot.1.e})}, we have
\begin{align*}
 \|\mathbb{G}_{\eps, F}I\|^{2}_{H^{s}_{x}L^{2}_{\xi}}&\leq\big(1+|\pi\eps k |^{2}\big)^{s}
\sum_{|\eps k|<\de}\sum_{j=0}^{4}|e^{\sigma_{j}(-\eps k)t}||\big<e_{j}(\eps k ), \hat{I}_{k}\big>_{\xi}|^{2} \\
 &\leq C
\sum_{|\eps k|<\de}\sum_{j=0}^{4}|e^{\sigma_{j}(-\eps k)t}|\| \hat{I}_{k}\|_{L^{2}_{\xi}}^{2}\\
&\leq C
\sum_{|\eps k|<\de}\sum_{j=0}^{4}e^{-a_{j,2}|k\eps|^{2}[1+O(\de^{2})]t}\| \hat{I}_{k}\|_{L^{2}_{\xi}}^{2}\\
&\leq C
e^{-O(1)\eps^{2}t}\| I\|_{L^{2}_{x}L^{2}_{\xi}}^{2}\,.
\end{align*}
\qed

$\mathbf{Remark}$
(i) If $\eps>\de$, we do not have long wave part, i.e. $\mathbb{G}^{t}_{\eps, L}=0$.

\noindent(ii) The high order estimate of short wave part requires regularity in $x$. This is because $|\pi\eps k|$ may not be bounded. One needs the regularity of $I$ to ensure the decay of $\mathbb{G}^{t}_{\eps, S}$ in time.

\noindent(iii) In order to remove the regularity assumption in $x$, we need the Picard-type iteration for constructing the increasingly regular kinetic-like waves in section \ref{se4}.
\begin{theorem}
For any $I\in H_{x}^{2}L_{\xi}^{2}$ and satisfies the zero mean conditions {\rm(\ref{bot.1.e})}, we have the following exponential time decay estimate about the linearized Boltzmann equation {\rm(\ref{bot.2.a})}
\begin{equation*}
\|\mathbb{G}_{\eps}^{t}I\|_{L^{\infty}_{x}L^{2}_{\xi}}\leq e^{-\la_{S}t}\|I\|_{H^{2}_{x}L^{2}_{\xi}}
+e^{-\la_{L}t}\|I\|_{L^{2}_{x}L^{2}_{\xi}}\,.
\end{equation*}
\noindent{\rm(i)} If $\eps>\de$, then $\la_{S}=O(1)$ and $\la_{L}=\infty$,
\\
\noindent{\rm(ii)} If $\eps\leq\de$, then $\la_{S}=O(1)$ and $\la_{L}=O(1)\eps^{2}$.
\end{theorem}
\section{Fluid Part}\label{se3}

In this section, we improve the estimate of the fluid part. Recall the fluid part of the Boltzmann equation (\ref{bot.2.g})
\begin{equation*}
\mathbb{G}^{t}_{\eps, F}I=\sum_{|\eps k|<\de}\sum_{j=0}^{4}e^{\sigma_{j}(-\eps k)t}e^{i\pi\eps k\cdot x}\big<e_{j}(\eps k ), \hat{I}_{k}\big>_{\xi}e_{j}(\eps k )\,.
\end{equation*}
We have
\begin{align*}
  \|\mathbb{G}_{\eps, F}^{t}I\|^{2}_{L^{2}_{\xi}}&=\sum_{j=0}^{4}\Big|\sum_{|\eps k|<\de}e^{\sigma_{j}(-\eps k)t}e^{i\pi\eps k\cdot x}\big<e_{j}(\eps k ), \hat{I}_{k}\big>_{\xi}\Big|^{2} \\
 &=\sum_{j=0}^{4}\Big|\frac{1}{|\T^{3}_{1/\eps}|}\sum_{|\eps k|<\de}\int_{\T^{3}_{1/\eps}}\int_{\R^{3}}e^{\sigma_{j}(-\eps k)t}e^{i\pi\eps k\cdot (x-y)}e_{j}(\eps k )I(y,\xi)d\xi dy\Big|^{2}\,.
\end{align*}
Note that
$$
\frac{1}{|\T^{3}_{1/\eps}|}\sum_{|\eps k|<\de}e^{-(\eps |k|)^{2}t}=\frac{1}{t^{3/2}}\frac{1}{|\T^{3}_{1/\eps}|}\sum_{|\eps k|<\de t^{1/2}}e^{-(\eps |k|)^{2}}
\to \frac{1}{t^{3/2}}\int_{B_{\de t^{1/2}}(0)}e^{-y^{2}}dy
$$
as $\eps\to 0$, this means for any $\alpha_{0}>0$ there exists $\de_{0}>0$ such that if $\eps<\de_{0}$, then
$$
\frac{1}{|\T^{3}_{1/\eps}|}\sum_{|\eps k|<\de t^{1/2}}e^{-(\eps |k|)^{2}}<\alpha_{0}+ \int_{\R^{3}}e^{-y^{2}}dy
\equiv C_{0}\,.
$$
Hence
\begin{align*}
  \|\mathbb{G}_{\eps, F}^{t}I\|_{L^{2}_{\xi}}&\leq
\Big(\frac{1}{|\T^{3}_{1/\eps}|}\sum_{|\eps k|<\de}e^{-(\eps |k|)^{2}t}\Big)\int_{\T^{3}_{1/\eps}}\|I\|_{L^{2}_{\xi}}dx \\
 &\leq O(1)
\Big(\frac{1}{|\T^{3}_{1/\eps}|}\sum_{|\eps k|<\de}e^{-(\eps |k|)^{2}t}\Big)e^{-\eps^{2}t}\\
&\leq O(1)
\frac{C_{0}}{(1+t)^{3/2}}e^{-\eps^{2}t}\,.
\end{align*}
\begin{theorem}
Assume that $\eps<\de$, $I\in L_{\xi}^{2}$ with compact support in $x$ and satisfies the zero mean conditions {\rm(\ref{bot.1.e})}, then there exist $C_{0}, \de_{0}>0$ such that if $0<\eps<\de_{0}$, then
\begin{equation*}
\displaystyle\|\mathbb{G}_{\eps, F}^{t}I\|_{L^{2}_{\xi}}\leq  O(1)
\frac{C_{0}}{(1+t)^{3/2}}e^{-\eps^{2}t} \,.
\end{equation*}
\end{theorem}
If $\eps\to 0$, i.e. the whole space case, the pointwise estimate of the fluid part becomes
$$
\|\mathbb{G}_{0, F}^{t}I\|_{L^{2}_{\xi}}\leq O(1)\frac{1}{(1+t)^{3/2}}\,.
$$
This recover the whole space result in \cite{[LiuYu]}.
\section{Kinetic Part and Remainder Part}\label{se4}

In this section, we will apply the kinetic decomposition and mixture lemma to construct the kinetic and remainder parts. We rewrite the linearized Boltzmann equation (\ref{bot.2.a}) as
\begin{equation}\label{bot.3.a}
\left\{\begin{array}{l} \displaystyle\pa_{t}f+\xi\cdot\nabla_{x}f+\nu(\xi)f-K_{s}f=K_{r}f\,,
\\ \\
\displaystyle f(x,0,\xi)=I(x,\xi)\,.
\\
\end{array}
\right.\end{equation}
Now, we design a Picard type iteration, which treat the regular part $K_{r}f$ as the source term. The $-1$ order approximation of the linearized Boltzmann equation (\ref{bot.3.a}) is the damped transport equation
\begin{equation}\label{bot.3.b}
\left\{\begin{array}{l} \pa_{t}h^{(-1)}+\xi\cdot\nabla_{x}h^{(-1)}+\nu(\xi)h^{(-1)}-K_{s}f^{(-1)}=0\,,
\\ \\
h^{(-1)}(x,0,\xi)=I(x,\xi)\,.
\\
\end{array}
\right.\end{equation}
The difference $f-h^{(-1)}$ satisfies the equation
\begin{equation*}
\left\{\begin{array}{l} \pa_{t}(f-h^{(-1)})+\xi\cdot\nabla_{x}(f-h^{(-1)})+\nu(\xi)(f-h^{(-1)})=K(f-h^{(-1)})+K_{r}h^{(-1)}\,,
\\ \\
(f-h^{(-1)})(x,0,\xi)=0\,.
\\
\end{array}
\right.\end{equation*}
Thus we can define the zero order approximation $h^{(0)}$
\begin{equation}\label{bot.3.c}
\left\{\begin{array}{l} \pa_{t}h^{(0)}+\xi\cdot\nabla_{x}h^{(0)}+\nu(\xi)h^{(0)}=K_{r}h^{(-1)}\,,
\\ \\
h^{(0)}(x,0,\xi)=0\,.
\\
\end{array}
\right.\end{equation}
In general, we can define the $j^{\hbox{th}}$ order approximation $h^{(j)}$, $j\geq 1$ as
\begin{equation}\label{bot.3.d}
\left\{\begin{array}{l} \pa_{t}h^{(j)}+\xi\cdot\nabla_{x}h^{(j)}+\nu(\xi)h^{(j)}=Kh^{(j-1)}\,,
\\ \\
h^{(j)}(x,0,\xi)=0\,.
\\
\end{array}
\right.\end{equation}
This means that the solution $f=\mathbb{G}_{\eps}^{t}I$ of the linearized Boltzmann equation (\ref{bot.2.a}) can be rewritten as a series
$$
f=h^{(-1)}+h^{(0)}+h^{(1)}+\cdot\cdot\cdot\,.
$$
Let $\mathbb{S}^{t}$ and $\mathbb{O}^{t}$ denote the solution operators of the following equations,
\begin{equation*}\label{bot.3.e}
\left\{\begin{array}{l} \pa_{t}g +\xi\cdot\nabla_{x}g +\nu(\xi)g =0\,,
\\ \\
g (x,0,\xi)=g_{0}(x,\xi)\,,
\end{array}
\right.
\end{equation*}
and
\begin{equation*}\label{bot.3.f}
\left\{\begin{array}{l} \pa_{t}j +\xi\cdot\nabla_{x}j +\nu(\xi)j -K_{s}j=0\,,
\\ \\
j (x,0,\xi)=j_{0}(x,\xi)\,,
\end{array}
\right.
\end{equation*}
i.e.
$$
g(x,t,\xi)=\mathbb{S}^{t}g_{0}(x,\xi)\,,\quad j(x,t,\xi)=\mathbb{O}^{t}j_{0}(x,\xi)\,.
$$
By standard energy estimate, maximum principle and property of the integral operator $K$ in (\ref{bot.1.h}), we have the following results about the operator $\mathbb{S}^{t}$ and $\mathbb{O}^{t}$.
\begin{lemma}\label{lem1}{\rm\cite{[LiuYu]}}
For any $\beta\geq 0$, we have
\begin{equation}\label{bgk.4.h}
\left\{\begin{array}{l}\|\mathbb{S}^{t}g_{0}\|_{L^{2}_{x}L^{2}_{\xi}}\leq e^{-\nu_{0}t}\|g_{0}\|_{L^{2}_{x}L^{2}_{\xi}}\,,
\\ \\
\|\mathbb{S}^{t}g_{0}\|_{L^{\infty}_{x}L^{\infty}_{\xi,\beta}}\leq e^{-\nu_{0}t}\|g_{0}\|_{L^{\infty}_{x}L^{\infty}_{\xi,\beta}}\,.
\end{array}
\right.
\end{equation}
and
\begin{equation}\label{bgk.4.h1}
\left\{\begin{array}{l}\|\mathbb{O}^{t}j_{0}\|_{L^{2}_{x}L^{2}_{\xi}}\leq e^{-\nu_{0}t/2}\|j_{0}\|_{L^{2}_{x}L^{2}_{\xi}}\,,
\\ \\
\|\mathbb{O}^{t}j_{0}\|_{L^{\infty}_{x}L^{\infty}_{\xi,\beta}}\leq e^{-\nu_{0}t/2}\|j_{0}\|_{L^{\infty}_{x}L^{\infty}_{\xi,\beta}}\,.
\end{array}
\right.
\end{equation}
\end{lemma}
The following lemma gives the $L^{2}_{x}L^{2}_{\xi}$ and $L^{\infty}_{x}L^{\infty}_{\xi,\beta}$ estimate of $h^{(j)}$.
\begin{lemma}\label{lem2}For $\beta\geq 0$, $j\geq -1$, we have
\begin{equation}\begin{array}{l}\label{bgk.4.j}
\|h^{(j)}\|_{L^{2}_{x}L^{2}_{\xi}}\leq t^{j+1}e^{-\nu_{0}t/2}\|I\|_{L^{2}_{x}L^{2}_{\xi}}\,,
\\ \\
\|h^{(j)}\|_{L^{\infty}_{x}L^{\infty}_{\xi,\beta}}\leq t^{j+1}e^{-\nu_{0}t/2}\|I\|_{L^{\infty}_{x}L^{\infty}_{\xi,\beta}}\,.
\end{array}
\end{equation}
\end{lemma}
\noindent{\it Proof.} We will prove by induction. The case $j=-1$ immediately follows from (\ref{bot.3.b}) and Lemma \ref{lem1}. For $j=0$, by the definition of $h^{(0)}$ in (\ref{bot.3.c}) and Duhamel principle,
\begin{equation*}\begin{array}{l}
\displaystyle h^{(0)}(x,t,\xi)=\int_{0}^{t}\mathbb{S}^{t-s_{1}}K_{r}\mathbb{O}^{s_{1}}I(\cdot,s_{1})ds_{1}
\,,
\end{array}
\end{equation*}
using proposition \ref{pro2} and lemma \ref{lem1}, we have
$$
\|h^{(0)}\|_{L^{2}_{x}L^{2}_{\xi}}\leq O(1) te^{-\nu_{0}t/2}\|I\|_{L^{2}_{x}L^{2}_{\xi}}\,.
$$
Assume it holds for $j$, then by (\ref{bot.1.g}), (\ref{bot.3.d}) and (\ref{bgk.4.h}), we have
\begin{align*}
  \|h^{(j+1)}\|_{L^{2}_{x}L^{2}_{\xi}}&=\Big\|\int_{0}^{t}\mathbb{S}^{t-s}(Kh^{(j)})(\cdot,s)ds\Big\|_{L^{2}_{x}L^{2}_{\xi}}\\
  &\leq \int_{0}^{t}e^{-\nu_{0}(t-s)}e^{-\nu_{0}s/2}s^{j+1}\|I\|_{L^{2}_{x}L^{2}_{\xi}}ds
  \\
  &\leq t^{j+2}e^{-\nu_{0}t/2}\|I\|_{L^{2}_{x}L^{2}_{\xi}}\,.
\end{align*}
The estimate of $L^{\infty}_{x}L^{\infty}_{\xi,\beta}$ norm is similar and hence we omit the detail.
\qed

Now, we can define the kinetic decomposition
\begin{equation}\label{bot.4.a}
\mathbb{G}_{\eps}^{t}I=\sum_{j=-1}^{4}h^{(j)}+\mathcal{R}\,,
\end{equation}
then the tail term $\mathcal{R}$ satisfies the equation
\begin{equation*}\label{bot.4.b}
\left\{\begin{array}{l}
\pa_{t}\mathcal{R}+\xi\cdot\nabla_{x} \mathcal{R}=L\mathcal{R}+Kh^{(4)}\,,
\\ \\
\mathcal{R}(x,0,\xi)=0\,.
\\
\end{array}
\right.\end{equation*}
Also, the kinetic part and the remainder part can be defined as follows:
\begin{equation*}\label{bot.4.c}
\mathbb{G}_{\eps, K}^{t}I=\sum_{j=-1}^{4}h^{(j)}\,, \quad \mathbb{G}_{\eps, R}^{t}I=\mathcal{R}-\mathbb{G}_{\eps, F}^{t}I\,.
\end{equation*}
Combining long wave short wave decomposition (\ref{bot.2.e}), (\ref{bot.2.g}) and kinetic decomposition (\ref{bot.4.a}), we have
\begin{equation*}\label{bot.4.d}
\mathbb{G}_{\eps, R}^{t}I=\mathbb{G}^{t}_{\eps, S}I+\mathbb{G}^{t}_{\eps, L;\perp}I-\sum_{j=-1}^{4}h^{(j)}=\mathcal{R}-\mathbb{G}_{\eps, F}^{t}I\,,
\end{equation*}
hence by (\ref{bot.2.f}), (\ref{bot.2.h}) and (\ref{bgk.4.j}),
\begin{equation*}\label{bot.4.e}
\big\|\mathbb{G}_{\eps, R}^{t}I\big\|_{L^{2}_{x}L^{2}_{\xi}}\leq t^{5}e^{-O(1)t}\|I\|_{L^{2}_{x}L^{2}_{\xi}}\,.
\end{equation*}
For the high order estimate, we have
\begin{equation}\label{bot.4.f}
\frac{d}{dt}\|\pa^{2}_{x}\mathbb{G}_{\eps, R}^{t}I\|_{L^{2}_{x}L^{2}_{\xi}}\leq-C\|\pa^{2}_{x}\mathbb{G}_{\eps, R}^{t}I\|_{L^{2}_{x}L^{2}_{\xi}}
+\|K(\pa^{2}_{x}h^{(4)})(\cdot,t)\|_{L^{2}_{x}L^{2}_{\xi}}\,.
\end{equation}
We only need to estimate
\begin{equation}\label{bot.4.g}
\int_{0}^{t}\|K(\pa^{2}_{x}h^{(4)})(\cdot,t-s)\|_{L^{2}_{x}L^{2}_{\xi}} ds=O(1)\,.
\end{equation}
To proceed, we define the $j^{th}$ Mixture operator as follow:
\begin{equation*}\label{bgk.4.r}
\mathbb{M}^{t}_{j}f_{0}=\int_{0}^{t}\int_{0}^{s_{1}}\cdot\cdot\cdot\int_{0}^{s_{2j-1}}\mathbb{S}^{t-s_{1}}K
\mathbb{S}^{s_{1}-s_{2}}K\mathbb{S}^{s_{2}-s_{3}}K\cdot\cdot\cdot\mathbb{S}^{s_{2j-1}-s_{2j}}K\mathbb{S}^{s_{2j}}
f_{0}ds_{2j}\cdot\cdot\cdot ds_{1}\,.
\end{equation*}
This form indicates that there are two essential mixing mechanisms:
\\
(i) The mixing mechanism in $x$ is due to particles traveling in different velocity $\xi$. This is represented by the operator $\mathbb{S}^{t}$.
\\
(ii) The mixing mechanism in $\xi$ is due to the compact operator $K$.
\\ \\
Under this definition, we have
\begin{equation}\label{bgk.4.s}
h^{(4)}(x,t,\xi)=\int_{0}^{t}\mathbb{M}^{t-s_{0}}_{2}K_{r}\mathbb{O}^{s_{0}}I(\cdot,s_{0})ds_{0}\,.
\end{equation}

\begin{lemma}\label{mix}{\rm(mixture lemma \cite{[Kuo], [LiuYu]})} For any $f_{0}\in L^{2}_{x}H^{j}_{\xi}$, $j=1,2$, we have
\begin{equation}\begin{array}{l}\label{bgk.4.t}
\|\pa^{j}_{x}\mathbb{M}^{t}_{j}f_{0}\|_{L^{2}_{x}L^{2}_{\xi}}\leq t^{j}e^{-2\nu_{0}t/3}\|f_{0}\|_{L^{2}_{x}H^{j}_{\xi}}\,.
\end{array}
\end{equation}
\end{lemma}
The mixture lemma states that the mixture of the two operators
$\mathbb{S}$ and $K$ in $\mathbb{M}^{t}_{j}$ transports the regularity in the microscopic velocity
$\xi$ to the regularity of the space time $(x,t)$. The proof of this lemma will be given in Appendix. We can estimate $(\ref{bot.4.g})$ by using the mixture lemma:
\begin{lemma}
\begin{equation}\label{bgk.4.u}
\int_{0}^{t}\|K(\pa^{2}_{x}h^{(4)})(\cdot,t-s)\|_{L^{2}_{x}L^{2}_{\xi}} ds=O(1)\|I\|_{L^{2}_{x}L^{2}_{\xi}}\,.
\end{equation}
\end{lemma}
\noindent{\it Proof.} By (\ref{bgk.4.s}) and (\ref{bgk.4.t}), we have
\begin{align}
 \|\pa^{2}_{x}h^{(4)}(\cdot,s)\|_{L^{2}_{x}L^{2}_{\xi}}&\leq\int_{0}^{s}(s-s_{0})^{2}e^{-2\nu_{0}(s-s_{0})/3}
\|\pa^{2}_{\xi}K_{r}\mathbb{O}^{s_{0}}I\|_{L^{2}_{x}L^{2}_{\xi}}ds_{0}\label{bgk.4.v} \\
&\leq e^{-\nu_{0}s/2}\int_{0}^{s}(s-s_{0})^{i}ds_{0}\|I\|_{L^{2}_{x}L^{2}_{\xi}}\nonumber\\
&\leq s^{3}e^{-\nu_{0}s/2}\|I\|_{L^{2}_{x}L^{2}_{\xi}}\nonumber\,,
\end{align}
moreover, (\ref{bot.1.g}) and (\ref{bgk.4.v}) imply
\begin{align*}
 \int_{0}^{t}\|K(\pa^{2}_{x}h^{(4)})(\cdot,t-s)\|_{L^{2}_{x}L^{2}_{\xi}} ds
&\leq\int_{0}^{t}
\|K\|_{L^{2}_{x}L^{2}_{\xi}}\|\pa^{2}_{x}h^{(4)}\|_{L^{2}_{x}L^{2}_{\xi}}ds\\
 &\leq\|I\|_{L^{2}_{x}L^{2}_{\xi}}\int_{0}^{t}(t-s)^{3}e^{-\nu_{0}(t-s)/2}ds
\\
&=O(1)\|I\|_{L^{2}_{x}L^{2}_{\xi}}\,.
\end{align*}
\qed

By (\ref{bot.4.f}) and (\ref{bgk.4.u}), we obtain
\begin{equation*}
\big\|\mathbb{G}_{\eps, R}^{t}I\big\|_{H^{2}_{x}L^{2}_{\xi}}\leq O(1)e^{-O(1)t}\,.
\end{equation*}
This completes the proof of the main theorem \ref{theorem1}.
\section{Appendix: Proof of the Mixture Lemma}
The goal of this appendix is to give a short and direct proof of the mixture lemma. This lemma was introduced by Liu-Yu \cite{[LiuYu], [LiuYu2], [LiuYu3], [LiuYu1]} and the proof relies on the explicit solution of the damped transport equation. In order to avoid constructing the explicit solution, we need to introduce a differential operator:
 $$
 \mathcal{D}_{t}=t\nabla_{x}+\nabla_{\xi}\,.
 $$
It is important that $\mathcal{D}_{t}$ commutes with free transport operator:
$$
[\mathcal{D}_{t},\pa_{t}+\xi\cdot\nabla_{x} ]=0\,.
$$
We have the following estimates about differential operator $\mathcal{D}_{t}$ and solution operator $\mathbb{S}^{t}$.
\begin{lemma}\label{lemma-a}
For any $f_{0}\in L^{2}_{x}H^{j}_{\xi}$, $j=1,2$, there exists $\eta_{0}$ small enough such that
\begin{equation}\label{ml.1}
\left\{\begin{array}{l}
\displaystyle\|\mathcal{D}_{t}^{j}\mathbb{S}^{t}f_{0}\|_{L^{2}_{x}L^{2}_{\xi}}\leq e^{-(\nu_{0}-\eta_{0})t}\|f_{0}\|_{L^{2}_{x}H^{j}_{\xi}}\,,
\\ \\
\displaystyle\|\mathcal{D}_{t}\mathbb{S}^{t}f_{0}-\mathbb{S}^{t}\nabla_{\xi}f_{0}\|_{L^{2}_{x}L^{2}_{\xi}}\leq e^{-(\nu_{0}-\eta_{0})t}\|f_{0}\|_{L^{2}_{x}L^{2}_{\xi}}\,,
\\ \\
\displaystyle\|\mathcal{D}_{t}^{2}\mathbb{S}^{t}f_{0}-\mathcal{D}_{t}\mathbb{S}^{t}\nabla_{\xi}f_{0}\|_{L^{2}_{x}L^{2}_{\xi}}\leq e^{-(\nu_{0}-\eta_{0})t}\|f_{0}\|_{L^{2}_{x}H^{1}_{\xi}}\,.
\\
\end{array}
\right.\end{equation}
\end{lemma}
\textbf{Remark:} Although the integral operator $K$ has smoothing property, it only allows us differential with respect to $\xi$ once, if we calculate second order Mixture operator $\partial^{2}_{x}\mathbb{M}^{t}_{2}f_{0}$, the second derivative term $\pa^{2}_{\xi}K(\xi, \xi_{*})$ appears, the cancelation properties $(\ref{ml.1})_{2}$ and $(\ref{ml.1})_{3}$ can overcome this difficulty.
\\ \\
\noindent{\it Proof.} We can check the commutator
\begin{equation}\begin{array}{l}\label{bot.6.a}
\displaystyle[\mathcal{D}_{t},\nu(\xi) ]h=\nabla_{\xi}\nu(\xi)h\,,
\\ \\
\displaystyle[\mathcal{D}^{2}_{t},\nu(\xi) ]h=2[\mathcal{D}_{t},\nu(\xi) ]\mathcal{D}_{t}h+\big[\mathcal{D}_{t},[\mathcal{D}_{t},\nu(\xi)] \big]h
=2\nabla_{\xi}\nu(\xi)\mathcal{D}_{t}h+\nabla_{\xi}^{2}\nu(\xi)h\,.
\end{array}\end{equation}
Let $f^{(j)}=\mathcal{D}^{j}_{t}\mathbb{S}^{t}f_{0}$, then the energy estimate for $f^{(0)}(=\mathbb{S}^{t}f_{0})$ gives
\begin{equation*}\begin{array}{l}
\displaystyle \|f^{(0)}\|^{2}_{L^{2}_{x}L^{2}_{\xi}}+\int_{0}^{t}
\|\nu^{1/2}(\xi)f^{(0)}\|^{2}_{L^{2}_{x}L^{2}_{\xi}}ds
\leq\|f_{0}\|^{2}_{L^{2}_{x}L^{2}_{\xi}}
\,.
\end{array}\end{equation*}
For $f^{(1)}(=\mathcal{D}_{t}\mathbb{S}^{t}f_{0})$, it solves the equation
\begin{equation}\label{bot.6.b}
\left\{\begin{array}{l}
\displaystyle\pa_{t}f^{(1)}+\xi\cdot\nabla_{x}f^{(1)}=-\nu(\xi) f^{(1)}-[\mathcal{D}_{t},\nu(\xi) ]f^{(0)}\,,
\\ \\
\displaystyle f^{(1)}(x,0,\xi)=\nabla_{\xi}f_{0}(x,\xi)\,.
\\
\end{array}
\right.\end{equation}
By (\ref{bot.1.f}), (\ref{bot.6.a}) and energy estimate, we have
\begin{equation*}\begin{array}{l}
\displaystyle\frac{1}{2}\frac{d}{dt}\|f^{(1)}\|^{2}_{L^{2}_{x}L^{2}_{\xi}}+\|\nu^{1/2}(\xi)f^{(1)}\|^{2}_{L^{2}_{x}L^{2}_{\xi}}
\leq \eta_{0}\|f^{(1)}\|^{2}_{L^{2}_{x}L^{2}_{\xi}}
+C\|f^{(0)}\|^{2}_{L^{2}_{x}L^{2}_{\xi}}\,,
\end{array}\end{equation*}
choose $\eta_{0}$ small enough, we obtain
$$
\|\mathcal{D}_{t}\mathbb{S}^{t}f_{0}\|_{L^{2}_{x}L^{2}_{\xi}}\leq e^{-(\nu_{0}-\eta_{0})t}\|f_{0}\|_{L^{2}_{x}H^{1}_{\xi}}\,.
$$
For $f^{(2)}(=\mathcal{D}^{2}_{t}\mathbb{S}^{t}f_{0})$, we have
\begin{equation}\label{bot.6.c}
\left\{\begin{array}{l}
\displaystyle\pa_{t}f^{(2)}+\xi\cdot\nabla_{x}f^{(2)}=\nu(\xi) f^{(2)}+2[\mathcal{D}_{t},\nu(\xi) ]f^{(1)}+\big[\mathcal{D}_{t},[\mathcal{D}_{t},\nu(\xi)] \big]f^{(0)}\,,
\\ \\
\displaystyle f^{(2)}(x,0,\xi)=\nabla^{2}_{\xi}f_{0}(x,\xi)\,.
\\
\end{array}
\right.\end{equation}
Similar argument can get our result and hence we omit the detail, this proves $(\ref{ml.1})_{1}$.
For $(\ref{ml.1})_{2}$ and $(\ref{ml.1})_{3}$, for simplicity of notation, we can define $g_{0}=\nabla_{\xi}f_{0}$ and $g^{(j)}=\mathcal{D}_{t}\mathbb{S}^{t}g_{0}$, then $g^{(0)}$
 and $g^{(1)}$ satisfy the following equations respectively
\begin{equation}\label{bot.6.d}
\left\{\begin{array}{l}
\pa_{t}g^{(0)}+\xi\cdot\nabla_{x}g^{(0)}=-\nu(\xi) g^{(0)}\,,
\\ \\
\displaystyle g^{(0)}(x,0,\xi)=\nabla_{\xi}f_{0}\,,
\\
\end{array}
\right.\end{equation}
and
\begin{equation}\label{bot.6.e}
\left\{\begin{array}{l}
\pa_{t}g^{(1)}+\xi\cdot\nabla_{x}g^{(1)}=-\nu(\xi) g^{(1)}-[\mathcal{D}_{t},\nu(\xi) ]g^{(0)}\,,
\\ \\
\displaystyle g^{(1)}(x,0,\xi)=\nabla^{2}_{\xi}f_{0}\,.
\\
\end{array}
\right.\end{equation}
We can define $u^{(j)}=f^{(j)}-g^{(j-1)}$, $j=1,2$. Then $u^{(1)}=\mathcal{D}_{t}\mathbb{S}^{t}f_{0}-\mathbb{S}^{t}\nabla_{\xi}f_{0}$
and $u^{(2)}=\mathcal{D}_{t}^{2}\mathbb{S}^{t}f_{0}-\mathcal{D}_{t}\mathbb{S}^{t}\nabla_{\xi}f_{0}$. For $(\ref{ml.1})_{2}$, by (\ref{bot.6.b}) and (\ref{bot.6.d}), we can calculate that
$u^{(1)}$ solves the equation
\begin{equation*}
\left\{\begin{array}{l}
\pa_{t}u^{(1)}+\xi\cdot\nabla_{x}u^{(1)}=-\nu(\xi) u^{(1)}-[\mathcal{D}_{t},\nu(\xi) ]f^{(0)}\,,
\\ \\
\displaystyle u^{(1)}(x,0,\xi)=0\,,
\\
\end{array}
\right.\end{equation*}
the energy estimate, (\ref{bot.6.a}) and (\ref{bot.1.f}) gives
$$
\|u^{(1)}\|^{2}_{L^{2}_{x}L^{2}_{\xi}}\leq -(\nu_{0}-\eta_{0})\int_{0}^{t}\|u^{(1)}\|^{2}_{L^{2}_{x}L^{2}_{\xi}}ds+\|f_{0}\|^{2}_{L^{2}_{x}L^{2}_{\xi}}\,,
$$
this complete the proof of $(\ref{ml.1})_{2}$. For $(\ref{ml.1})_{3}$, similarly, by (\ref{bot.6.c}) and (\ref{bot.6.e}), we can calculate that $u^{(2)}$ solves the equation
\begin{equation*}
\left\{\begin{array}{l}
\pa_{t}u^{(2)}+\xi\cdot\nabla_{x}u^{(2)}-\nu(\xi) u^{(2)}
=2[\mathcal{D}_{t},\nu(\xi) ]f^{(1)}+\big[\mathcal{D}_{t},[\mathcal{D}_{t},\nu(\xi)] \big]f^{(0)}-[\mathcal{D}_{t},\nu(\xi) ]g^{(0)}\,,
\\ \\
\displaystyle u^{(2)}(x,0,\xi)=0\,,
\\
\end{array}
\right.\end{equation*}
similar argument can get our result.
\qed
\\ \\
\noindent{\it Proof of the Mixture Lemma.} For $j=1$, we can write down $\pa^{1}_{x}\mathbb{M}^{t}_{1}f_{0}$ as follows:
\begin{align*}
  \pa^{1}_{x}\mathbb{M}^{t}_{1}f_{0}(x,\xi)&=
  \int_{0}^{t}\int_{0}^{s_{1}}
\pa_{x}\mathbb{S}^{t-s_{1}}K
\mathbb{S}^{s_{1}-s_{2}}K\mathbb{S}^{s_{2}}f_{0}ds_{2}ds_{1}\\
&=\int_{0}^{t}\int_{0}^{s_{1}}\int_{\R^{6}}\pa_{x}\mathbb{S}^{t-s_{1}}W(\xi,\xi_{1})\mathbb{S}^{s_{1}-s_{2}}W(\xi_{1},\xi_{2})
\mathbb{S}^{s_{2}}f_{0}d\xi_{2}d\xi_{1}ds_{2}ds_{1}\,.
\end{align*}
Note that $[\pa_{x}, \mathbb{S}^{t}]=0$, $[\pa_{x},W]=0$, we can change the order of $(\pa_{x}, \mathbb{S}^{t})$ and $(\pa_{x},W)$. In order to get time integrability, we can rewrite $\pa^{1}_{x}\mathbb{M}^{t}_{1}f_{0}(x,\xi)$ as
\begin{align*}
\pa^{1}_{x}\mathbb{M}^{t}_{1}f_{0}(x,\xi)&=\int_{0}^{t}\int_{0}^{s_{1}}\int_{\R^{6}}\frac{s_{1}-s_{2}}{s_{1}}\mathbb{S}^{t-s_{1}}W(\xi,\xi_{1})\pa_{x}\mathbb{S}^{s_{1}-s_{2}}W(\xi_{1},\xi_{2})
\mathbb{S}^{s_{2}}f_{0}d\xi_{2}d\xi_{1}ds_{2}ds_{1}\\
  &+\int_{0}^{t}\int_{0}^{s_{1}}\int_{\R^{6}}\frac{s_{2}}{s_{1}}\mathbb{S}^{t-s_{1}}W(\xi,\xi_{1})\mathbb{S}^{s_{1}-s_{2}}W(\xi_{1},\xi_{2})
\pa_{x}\mathbb{S}^{s_{2}}f_{0}d\xi_{2}d\xi_{1}ds_{2}ds_{1}\,.
\end{align*}
Using the fact $t\nabla_{x}=\mathcal{D}_{t}-\nabla_{\xi}$, we have
\begin{align*}
 \pa^{1}_{x}\mathbb{M}^{t}_{1}f_{0}(x,\xi)&=\int_{0}^{t}\int_{0}^{s_{1}}\int_{\R^{6}}\frac{1}{s_{1}}\mathbb{S}^{t-s_{1}}W(\xi,\xi_{1})
\big(\mathcal{D}_{s_{1}-s_{2}}-\nabla_{\xi_{1}}\big)\mathbb{S}^{s_{1}-s_{2}}W(\xi_{1},\xi_{2})
\mathbb{S}^{s_{2}}f_{0}d\xi_{2}d\xi_{1}ds_{2}ds_{1} \\
&+\int_{0}^{t}\int_{0}^{s_{1}}\int_{\R^{6}}\frac{1}{s_{1}}\mathbb{S}^{t-s_{1}}W(\xi,\xi_{1})\mathbb{S}^{s_{1}-s_{2}}W(\xi_{1},\xi_{2})
\big(\mathcal{D}_{s_{2}}-\nabla_{\xi_{2}}\big)\mathbb{S}^{s_{2}}f_{0}d\xi_{2}d\xi_{1}ds_{2}ds_{1}\,.
\end{align*}
By $(\ref{ml.1})_{1}$ and integration by parts, we have
\begin{align*}
 \|\partial^{1}_{x}\mathbb{M}^{t}_{1}f_{0}\|_{L^{2}_{x}L^{2}_{\xi}}&\leq e^{-2\nu_{0}t/3}\big(\|f_{0}\|_{L^{2}_{x}L^{2}_{\xi}}
+\|\partial_{\xi}^{1}f_{0}\|_{L^{2}_{x}L^{2}_{\xi}}\big)\Big(\int_{0}^{t}\int_{0}^{s_{1}}\frac{1}{s_{1}}ds_{2}ds_{1}\Big) \\
&=te^{-2\nu_{0}t/3}\|f_{0}\|_{L^{2}_{x}H^{1}_{\xi}}\,.
\end{align*}
This proves the case $j=1$. For $j=2$, we can write down $\partial^{2}_{x}\mathbb{M}^{t}_{2}f_{0}$ as:
\begin{align*}
  &\phantom{xx}{} \partial^{2}_{x}\mathbb{M}^{t}_{2}f_{0}(x,\xi)\\
  &=\int_{\T}\int_{\R^{6\times 2}}\frac{s_{1}s_{3}}{s_{1}s_{3}}\partial^{2}_{x}
\Big[\mathbb{S}^{t-s_{1}}W(\xi,\xi_{1})
\mathbb{S}^{s_{1}-s_{2}}W(\xi_{1},\xi_{2})
\mathbb{S}^{s_{2}-s_{3}}W(\xi_{2},\xi_{3})\mathbb{S}^{s_{3}-s_{4}}W(\xi_{3},\xi_{4})
\mathbb{S}^{s_{4}}f_{0}\Big]d\Xi dS\,,
\end{align*}
where
$$
dS=ds_{1}\cdot\cdot\cdot ds_{4},\quad d\Xi=d\xi_{1}\cdot\cdot\cdot d\xi_{4}\,,\quad \T=[0,t]\times[0,s_{1}]\times\cdot\cdot\cdot\times[0,s_{3}]\,.
$$
In order to get time integrability, we need to decompose $s_{1}s_{3}$ as
$$s_{1}s_{3}=[(s_{1}-s_{2})+(s_{2}-s_{3})+(s_{3}-s_{4})+s_{4}][(s_{3}-s_{4})+s_{4}]\,.$$
We then have
\begin{equation*}\begin{array}{l}
\displaystyle\partial^{2}_{x}\mathbb{M}^{t}_{2}f_{0}(x,\xi)=\int_{\T}\int_{\R^{6\times 2}}\frac{1}{s_{1}s_{3}}(J_{1}f_{0}+J_{2}f_{0})d\Xi dS\,,
\end{array}\end{equation*}
where $J_{1}$ collects all the terms that each $W$ differential with respect to $\xi$ at most once, and $J_{2}$ collects all the terms that one of $W$ differential with respect to $\xi$ twice. More precisely,
\begin{align}
 J_{2}f_{0}
&=\mathbb{S}^{t-s_{1}}W(\xi,\xi_{1})
\mathbb{S}^{s_{1}-s_{2}}W(\xi_{1},\xi_{2})
\big[\mathcal{D}_{s_{2}-s_{3}}\mathbb{S}^{s_{2}-s_{3}}\big]\big[\nabla_{\xi_{3}}W(\xi_{2},\xi_{3})\big]\mathbb{S}^{s_{3}-s_{4}}W(\xi_{3},\xi_{4})
\mathbb{S}^{s_{4}}f_{0}\nonumber
\\
& +\mathbb{S}^{t-s_{1}}W(\xi,\xi_{1})
\mathbb{S}^{s_{1}-s_{2}}W(\xi_{1},\xi_{2})
\mathbb{S}^{s_{2}-s_{3}}\big[\nabla^{2}_{\xi_{3}}W(\xi_{2},\xi_{3})\big]\mathbb{S}^{s_{3}-s_{4}}W(\xi_{3},\xi_{4})
\mathbb{S}^{s_{4}}f_{0}\nonumber
\\
&+\mathbb{S}^{t-s_{1}}W(\xi,\xi_{1})
\mathbb{S}^{s_{1}-s_{2}}W(\xi_{1},\xi_{2})
\mathbb{S}^{s_{2}-s_{3}}W(\xi_{2},\xi_{3})\big[\mathcal{D}_{s_{3}-s_{4}}\mathbb{S}^{s_{3}-s_{4}}\big]\big[\nabla_{\xi_{4}}W(\xi_{3},\xi_{4})\big]
\mathbb{S}^{s_{4}}f_{0}\label{bot.6.f}
\\
&+\mathbb{S}^{t-s_{1}}W(\xi,\xi_{1})
\mathbb{S}^{s_{1}-s_{2}}W(\xi_{1},\xi_{2})
\mathbb{S}^{s_{2}-s_{3}}W(\xi_{2},\xi_{3})\mathbb{S}^{s_{3}-s_{4}}\big[\nabla^{2}_{\xi_{4}}W(\xi_{3},\xi_{4})\big]
\mathbb{S}^{s_{4}}f_{0}\nonumber
\\
&+\mathbb{S}^{t-s_{1}}W(\xi,\xi_{1})
\mathbb{S}^{s_{1}-s_{2}}W(\xi_{1},\xi_{2})
\mathbb{S}^{s_{2}-s_{3}}W(\xi_{2},\xi_{3})\big[\mathcal{D}^{2}_{s_{3}-s_{4}}\mathbb{S}^{s_{3}-s_{4}}\big]W(\xi_{3},\xi_{4})
\mathbb{S}^{s_{4}}f_{0}\nonumber
\\
&+\mathbb{S}^{t-s_{1}}W(\xi,\xi_{1})
\mathbb{S}^{s_{1}-s_{2}}W(\xi_{1},\xi_{2})
\mathbb{S}^{s_{2}-s_{3}}W(\xi_{2},\xi_{3})\big[\mathcal{D}_{s_{3}-s_{4}}\mathbb{S}^{s_{3}-s_{4}}\big]\big[\nabla_{\xi_{4}}W(\xi_{3},\xi_{4})\big]
\mathbb{S}^{s_{4}}f_{0}\nonumber\,.
\end{align}
The estimate of $J_{1}$ is similar to $j=1$,
$$
\Big\|\int_{\R^{6\times 2}}J_{1}f_{0}d\Xi\Big\|_{L^{2}_{x}L^{2}_{\xi}}\leq e^{-2\nu_{0}t/3}\|f_{0}\|_{L^{2}_{x}H^{2}_{\xi}}\,.
$$
For $J_{2}$, the first four terms of (\ref{bot.6.f}) can be estimated by $(\ref{ml.1})_{2}$ and the last two terms of (\ref{bot.6.f}) can be estimated by $(\ref{ml.1})_{3}$, then
$$
\Big\|\int_{\R^{6\times 2}}J_{2}f_{0}d\Xi\Big\|_{L^{2}_{x}L^{2}_{\xi}}\leq e^{-2\nu_{0}t/3}\|f_{0}\|_{L^{2}_{x}H^{2}_{\xi}}\,.
$$
This means
\begin{align*}
\|\pa^{2}_{x}\mathbb{M}^{t}_{2}f_{0}\|_{L^{2}_{x}L^{2}_{\xi}}\leq e^{-2\nu_{0}t/3}\|f_{0}\|_{L^{2}_{x}H^{2}_{\xi}}
\Big(\int_{\mathbb{T}}\frac{1}{s_{1}s_{3}}dS\Big)\leq t^{2}e^{-2\nu_{0}t/3}\|f_{0}\|_{L^{2}_{x}H^{2}_{\xi}}\,,
\end{align*}
this completes the proof of the lemma.
\qed

\end{document}